\documentclass[12pt,aps,prb,preprint, english]{revtex4}
\usepackage{graphicx}
\usepackage{dcolumn}
\usepackage{amsmath}
\usepackage{amssymb}
\usepackage{times}
\usepackage{babel}
\usepackage{epstopdf}
\usepackage{color}
\usepackage{float}
\usepackage{epstopdf}

\def\rv{{\bf r}}

\def\beq{\begin{equation}}
\def\eeq{\end{equation}}

\begin{document}
\title{$s$-wave scattering and the zero-range limit of the finite square well  in arbitrary dimensions}
\author{Aaron Farrell}
\affiliation{Department of Physics, St. Francis Xavier University, 
Antigonish, NS, Canada B2G 2W5} 
\author{Brandon P. van Zyl}
\affiliation{Department of Physics, St. Francis Xavier University, 
Antigonish, NS, Canada B2G 2W5}

\date{\today}

\begin{abstract}
We examine the zero-range limit of the finite square well in arbitrary dimensions through a systematic analysis of the reduced, $s$-wave two-body time-independent Schr\"odinger equation.  
A natural consequence  of our investigation is the requirement of a delta-function multiplied by a regularization operator to model the zero-range limit of the finite-square well when the dimensionality is
greater than one.
The case of two dimensions turns out to be surprisingly subtle,
and needs to be treated separately from all other dimensions.
\end{abstract}

\maketitle
\section{Introduction}
Every undergraduate student of physics familiarizes themselves with the mathematical machinery of quantum mechanics by solving the one-body, one-dimensional (1D) time-independent 
Schr\"odinger equation (TISE) for a variety of potentials which admit an exact solution.   One need only glance at any number of introductory quantum mechanics textbooks\cite{griffiths} to note that the attractive finite square well 
(FSW) and delta function potentials are among such standard potentials.  These potentials provide a pedagogical introduction to
the concepts of bound and
scattering states in 1D, while illustrating some of the pathologies that can occur when the potential is singular at a point.

Owing to the nature of the boundary conditions imposed on the wave function by the one-body potential, the mathematical analysis required for the solution 
of the bound and scattering states of the 1D FSW and delta-potential are quite different.  However, 
many textbooks make a point of commenting that the 1D delta potential results can be
recovered from the 1D FSW  of depth $V_0$ and range $b$, in the limits $V_0 \to \infty$, and $b \to 0$, provided that the area under the well, $V_0 b$, remain  constant
(see {\it e.g.,} problem 2.31 in Ref.~[\onlinecite{griffiths}]).  A question then naturally arises:  can higher dimensional zero-range interactions be similarly
constructed by taking an analogous limiting procedure?

Zero-range interactions have received renewed attention in the context of harmonically trapped cold atoms,~\cite{stoferle} where the dilute nature of the gas and ultra-low temperatures 
permit a treatment of the interactions between the charge neutral atoms as being primarily two-body and of zero-range, in the relative $s$-state.  
The symmetry of the relative $s$-state allows for a reduction of the two-body problem to an effective one-body problem that is already familiar to
undergraduate students exposed to elementary scattering theory.
Given that the dimensionality
of these systems can be manipulated in the laboratory,~\cite{lowdimensional} theoretical studies of such interacting, ultra-cold atomic systems involve 
the use of zero-range interactions appropriately generalized to two and three dimensions (2D and 3D, respectively).   

In this paper, we present a systematic extension of the analysis in 1D for obtaining a zero-range interaction from a FSW to arbitrary dimensions.  Our goal is
to illustrate, using only elementary quantum mechanics, how zero-range interactions in higher dimensions are constructed by starting from a well-studied finite range potential.  
We will focus solely on  two-body interactions in the relative $s$-state ($l=0$), with a view to readers interested in applications to, {\it e.g.,} modern ultra-cold atoms research.
We point out that the naive replacement of a finite square well in the $b\to 0$ limit with a {\em bare} delta-potential is only valid in 1D, whereas in higher dimensions, the
bare delta-potential must be ``regularized'' to avoid mathematical divergencies.  Our regularization of the delta-potential provides an intuitive introduction to the notion 
of pseudo-potentials in the context of zero-range interactions without having to invoke the more technical language of Green's functions and self-adjoint extensions 
that are sometimes used in the literature.  

The organization of our paper
is as follows.  In the next section, a brief introduction the hyper-spherical coordinates is presented, which then sets the stage for our general analysis of the $d$-dimensional 
FSW in Sec.~III.  In Sec.~IV, we examine the $b\to 0$ limit of the FSW in arbitrary dimensions.  In Sec.~V, we present our concluding remarks.

\section{Hyper-spherical Coordinates}
In the following analysis, we make use of  $d$-dimensional hyper-spherical coordinates.~\cite{ikhdair}  
These coordinates are comprised of a radius, $r$, and $d-1$ angular coordinates. 
In 3D, for example, the coordinates would be 
$r$ and the two angular variables $\theta$ and $\phi$.  
In what follows, we consider only the relative coordinate ({\it i.e.,} the centre-of-mass, 
${\bf R} \equiv ( {\bf r} + {\bf r}')/2$,
has already been removed so that we have an effectively one-body problem), and as such, 
the $d$-dimensional hyper-radius is defined as  
\begin{equation}\label{radius}
r \equiv |{\bf r} - {\bf r}'| =\sqrt{\sum_{i=1}^d (x_i - x_i')^2}
\end{equation} 
where $x_i$ is the $i$-th component of the first particles position, ${\bf r}$.   Owing to our focus on the relative $s$-state two-body interaction, the two-body wave function of the system will be hyper-spherically symmetric. 
It is useful to note that in this case, the integration of a function $f(r)$ over a hyper-sphere of radius $R$ is
given by
\begin{equation}\label{hyper}
\int f(r) d\tau = \frac{d \pi^{d/2}}{\Gamma(d/2+1)} \int_0^{R} f(r) r^{d-1}dr~,
\end{equation}
where the pre-factor is the $d$-dimensional analogue of the familiar 3D ``$4\pi$'' result. The use of the hyper-radius is straightforward in all dimensions except perhaps 1D, where
some confusion may arise.   Following the definition in Eq. (\ref{radius}) the 1D hyper-radius is
\begin{equation}
r_{1D} = \sqrt{(x - x')^2} =  |x|.
\end{equation}
Thus, while $r \in (0,\infty)$,  the Cartesian variable $x \in (-\infty,\infty)$.
Finally, note that in 1D the pre-factor of the integral given by Eq.~(2) is $2$, which reflects the fact that the analogous 1D $s$-state  function is {\em even} so that
\begin{equation}
\int_{-R}^{R} f(x)dx = 2\int_{0}^{R} f(x)dx =   2\int_{0}^{R} f(r)dr~.
\end{equation}
\section{The Finite Square Well in arbitrary dimensions}

We now consider  the $d$-dimensional generalization of the FSW, which takes the form
\begin{equation}
V(r) = - V_0\Theta(b-r)
\end{equation}
where $V_0>0$ is the strength of the interaction and $b$ is the range.  The $d$-dimensional TISE for $l=0$ then reads
\begin{equation}\label{inside}
-\frac{\hbar^2}{M} \nabla^2_d\psi_{<}(r) -V_0\psi_{<}(r) = E\psi_{<}(r) \ \ \ \ \ \ \ \ \ \ \ \ \ \ \ \ r<b
\end{equation}
and \begin{equation}\label{outside}
-\frac{\hbar^2}{M} \nabla^2_d\psi_{>}(r) = E\psi_{>}(r) \ \ \ \ \ \ \ \ \ \ \ \ \ \ \ \ r>b
\end{equation}
where $M$ is the mass of each particle,  $\psi_{<}(r)$ and $\psi_{>}(r)$ denote interior and exterior solutions, respectively, and $\nabla^2_d = \frac{d^2}{dr^2}+\frac{d-1}{r}\frac{d}{dr}$. 

\subsection{Scattering States: $E > 0$}

\subsubsection{Interior solution: $r<b$}
Let us first consider the region $r<b$ (the so-called interaction region). With $\eta^2 \equiv M(V_0+E)/\hbar^2= MV_0/\hbar^2 +k^2$ we can write Eq. (\ref{inside}) as
 \begin{equation}
\frac{d^2\psi_{<}(r)}{dr^2}+\frac{d-1}{r}\frac{d\psi_{<}(r)}{dr}= -\eta^2\psi_{<}(r).
\end{equation}  
This ordinary differential equation (ODE) is solved~\cite{handbook, note1} by a linear combination of Bessel functions of the first, $J_{\alpha}$, and second, $Y_{\alpha}$, kind with $\alpha\equiv d/2-1$.   However, since the $r<b$ region includes the origin, we discard $Y_{\alpha}$ as a possible solution as it diverges at the origin.  We then write our wave function in the interior region as
\begin{equation}\label{scatint}
\psi_{<}(r) = \frac{c_1\eta b^{d/2}}{J_{d/2}(\eta b)} \frac{ J_{\alpha}(\eta r)}{r^{d/2-1}}~,
\end{equation}
where $c_1$ is a constant of integration.
\subsubsection{Exterior solution: $r>b$}
Next, we consider the asymptotically free exterior region ($r>b$), where the TISE is given by  
 \begin{equation}\label{rgtb}
\frac{d^2\psi_{>}(r)}{dr^2}+\frac{d-1}{r}\frac{d\psi_{>}(r)}{dr}= -k^2\psi_{>}(r)~,
\end{equation} 
with $k^2\equiv ME/\hbar^2$.  
The solution to the above ODE can be written as~\cite{handbook}
\begin{equation}\label{scat}
\psi_{>}(r) = \frac{g_1J_{\alpha}(kr)+g_2Y_{\alpha}(kr)}{r^{d/2-1}},
\end{equation}  
where $g_1$ and $g_2$ are constants of integration.  Note that here, we retain the $Y_{\alpha}$ solution, in contrast to the interior region where it had to be ignored.

\subsubsection{Boundary conditions at $r=b$}
Given that our FSW potential has no singular behaviour, we require that the logarithmic derivative of the wave function be continuous at the boundary $r=b$
\begin{equation}
\left(\frac{\psi_{<}'(r)}{\psi_{<}(r)}\right)_{r=b}= \left(\frac{\psi_{>}'(r)}{\psi_{>}(r)}\right)_{r=b}~~,
\end{equation}
where $'\equiv d/dr$.

The result of Eq.~(12) is
\begin{equation}
k \left(\frac{g_1J_{\alpha+1}(kb)+g_2Y_{\alpha+1}(kb)}{g_1J_{\alpha}(kb)+g_2Y_{\alpha}(kb)}\right) = \eta \frac{J_{\alpha+1}(\eta b)}{J_{\alpha}(\eta b)},
\end{equation}
which allows us to write the ratio $\frac{g_2}{g_1}$ as
\begin{equation}\label{boundary}
\frac{g_2}{g_1} = \frac{ k J_{\alpha+1}(kb)J_{\alpha}(\eta b) - \eta J_{\alpha}(kb)J_{\alpha+1}(\eta b)}{\eta J_{\alpha+1}(\eta b)Y_{\alpha}(kb)-kY_{\alpha+1}(kb)J_{\alpha}(\eta b)}~.
\end{equation}
We may then re-write Eq.~(11) in the form familar from 3D scattering theory,~\cite{taylor} {\it viz.,}
\begin{equation}
\psi_{>}(r) = A\frac{\left(J_{\alpha}(kr) - \tan{\delta_0}Y_{\alpha}(kr)\right)}{r^{d/2-1}},
\end{equation}
where $A$ is a constant which is generally dimensionally dependent, and $\delta_0$ is called the $s$-wave {\em phase shift}.~\cite{taylor}  The $s$-wave phase shift characterizes the strength
of the scattering in the $l=0$ partial wave by the potential $V(r)$, at the energy $E=\hbar^2 k^2/M$.  Explicitly, we have
\begin{equation}
\tan(\delta_0) = 
\frac{ \eta J_{\alpha}(kb)J_{\alpha+1}(\eta b)-     k J_{\alpha+1}(kb)J_{\alpha}(\eta b)}{\eta J_{\alpha+1}(\eta b)Y_{\alpha}(kb)-kY_{\alpha+1}(kb)J_{\alpha}(\eta b)}~.
\end{equation}

Now, as $k \to 0$, the short-distance physics should  become irrelevant, and the scattering should not depend on the details of the short-range
two-body potential.   With this in mind, we
expand Eq.~(16) in the  $k \to 0$ limit, retaining only the leading order term in $k$,~\cite{note2} {\it viz.,}
\begin{equation}
-\tan(\delta_0) \simeq \frac{\pi}{\Gamma(d/2)\Gamma(d/2-1)}\left(\frac{k}{2}\right)^{d-2} \frac{1}{b^{2-d}\left(1+\frac{2-d}{\eta b}\frac{J_{\alpha}(\eta b)}{J_{\alpha+1}(\eta b)}\right)}~,~~~ d \ne 2
\end{equation}
where now, $\eta = \sqrt{MV_0/\hbar^2}$.~\cite{note2} 
We observe that for $d \ne 2$, we may  define the following quantity,
\begin{equation}
a^{d-2} \equiv -\frac{2^{d-2}}{\pi}\frac{1}{\Gamma(d/2)\Gamma(d/2-1)}\lim_{k\to 0} \frac{\tan(\delta_0)}{k^{d-2}}=
 \frac{1}{b^{2-d}\left(1+\frac{2-d}{\eta b}\frac{J_{\alpha}(\eta b)}{J_{\alpha+1}(\eta b)}\right)}~,
\end{equation}
from which we may re-write Eq.~(17) as
\begin{equation}
-\tan(\delta_0)  = \frac{\pi}{\Gamma(d/2)\Gamma(d/2-1)}\left(\frac{ka}{2}\right)^{d-2}~. 
\end{equation}
The crucial point here is that the $s$-wave phase shift, Eq.~(19), now depends on only a single quantity, $a$, and is {\em independent of the shape} of the potential.  
The quantity $a$ then completely characterizes the
$k \to 0$ limit of the scattering, so that {\em any}  short-range potential having the same value of $a$ will lead to absolutely identical scattering.  The single parameter
$a$ is called the $s$-wave scattering length, which is now generalized to $d \ne 2$ by Equation (18).
 It is easy to see that if we put $d=3$ in Eq.~(17), and apply Eq.~(18),  we obtain an explicit connection between the scattering length and the range, $b$, of the FSW
\begin{equation}
a = b\left(1 - \frac{\tan(\sqrt{MV_0/\hbar^2}b)}{\sqrt{MV_0/\hbar^2}b}\right)~,
\end{equation}
which is well-known for the 3D finite square well.  For completeness, it's also worthwhile noting that for $d=1$, we obtain 
\begin{equation}
a = b\left( 1+ \frac{ \cot(\sqrt{MV_0/\hbar^2}b)}{\sqrt{MV_0/\hbar^2}b}\right)~,
\end{equation}
which agrees with the symmetric 1D result discussed in great detail in Ref.~[\onlinecite{barlette}].~\cite{note3}

 The expression analogous to Eq.~(17) for $d=2$ is given by
\begin{equation}
-\tan{\delta_0} \simeq -\frac{\pi}{2\left(\ln{\frac{k}{2}}+\gamma+\ln{b}+ \frac{J_0(\eta b)}{\eta b J_1(\eta b)}\right)}~,~~~d=2~,
\end{equation}
where $\gamma \approx .577215665...$ is the Euler constant.
In this case, the 2D $s$-wave scattering length is defined by
\begin{equation}
a \equiv \lim_{k\to0} \frac{2e^{\frac{\pi}{2 \tan{\delta_0}}-\gamma}}{k}= be^{\frac{J_0(\eta b)}{\eta b J_1(\eta b)}}~,
\end{equation}
and Eq.~(22) becomes
\begin{equation}
-\tan{\delta_0} \simeq -\frac{\pi}{2\left(\ln{\frac{ka}{2}}+\gamma\right)}~,
\end{equation}
again illustrating that in the $k \to 0$ limit, the scattering is completely characterized by the $s$-wave scattering length, $a$.

The rather formal definition of the scattering length, {\it viz.,} Eqs.~(18) and (23), for $d \ne 2$ and $d=2$, respsectively, can be given a more
familiar geometric interpretation as follows.  Let us first recall the low energy behaviour of $J_{\alpha}$ and $Y_{\alpha}$:\cite{handbook}

\begin{equation}\label{SmallJ}
J_{\alpha}(kr) \to \frac{1}{\Gamma(d/2)}\left(\frac{kr}{2}\right)^{d/2-1}~,
\end{equation} 
and
 \begin{equation}\label{SmallY}
Y_{\alpha}(kr) = \left\{
     \begin{array}{lr}
       \frac{2}{\pi} \left( \ln{\left(\frac{kr}{2}\right)} + \gamma \right) &  d = 2\\
      -\frac{\Gamma(d/2-1)}{\pi}\left(\frac{2}{kr}\right)^{d/2-1}  & d \ne 2~,\\
     \end{array}
   \right.
\end{equation}
from which we obtain the $k \to 0$ scattering solutions, {\it viz.,}
\begin{equation}
\psi_{>}(r) \sim \left(1+\frac{\tan{\delta_0}\Gamma(d/2-1)\Gamma(d/2)}{\pi} \left(\frac{2}{kr}\right)^{d-2}\right)~,~~~~d\ne 2~,
\end{equation}
and 
\begin{equation}
\psi_{>}(r) \sim \left(1-\frac{2\tan{\delta_0}}{\pi}\left(\ln{\frac{kr}{2}}+\gamma\right)\right)~,~~~~d=2~.
\end{equation}
Inserting Eq.~(19) into Eq.~(27) and Eq.~(24) into Eq.~(28), we obtain (to within an overall unimportant constant)
\begin{equation}\label{seriesss}
\psi_{>}(r) \sim 1 -  \frac{a^{d-2}}{r^{d-2}} ~,~~~~d \ne 2~,
\end{equation}
and
\begin{equation}\label{seriesss2d}
\psi_{>}(r) \sim 1 -  \frac{\ln{kr/2}+\gamma}{\ln{ka/2}+\gamma}~,~~~~d=2~.
\end{equation} 
  
Let us first discuss $a>0$ (attractive potential).  In this case, the $s$-wave scattering length is the value at which the $k \to 0$ scattering wave function obtains
a node (for $r>b$).  Thus, the low-energy scattering wave function is ``bent over'' so that it has a negative slope at the boundary, implying that it can be joined to an exponentially decaying 
 solution corresponding
to a genuine bound state.  Therefore, the $k \to 0$ {\em scattering state} can be connected with a shallow {\em bound state} when $a>0$.  
Conversely, $a<0$ (repulsive potential), indicates the absence of a bound state.  Indeed, it is the backward extrapolation of the
$k\to 0$ wave function to $r<0$ which provides a measure of the negative scattering length.  An excellent, and more thorough,  discussion of the interpretation of the $s$-wave scattering length and its
sign can be found in Ref.~[\onlinecite{ahmed}].

It is interesting to note that in 2D, the scattering length, Eq.~(23), is {\em always} positive.
Within this context, note that in 1D and 2D, $a>0$ no matter how small the depth, $V_0$, of the well.  This observation leads to the ``well-known'' conclusion that in 1D and
2D, an {\em arbitrarily weak} attractive short range interaction will {\em always} support at least one bound state.~\cite{chandan}  Loosely speaking, in 1D and 2D, any depth, $V_0$, is enough
to bend the wave function so that $a>0$, whereas in 3D, a {\em minimum} depth is required before the first bound state occurs; for the 3D FSW, the minimum depth is readily seen to
be $V_0 > \pi^2\hbar^2/4Mb^2$.~\cite{shea2008}   In the very special case of 2D, where the scattering length is always positive (or $a\to \infty$ when $J_1$ in  Eq.~(23) has a zero), there are only
bound and loosely bound states~\cite{ahmed} for an attractive short-range potential.  Indeed, $a \to \infty$ signals the appearance of the next loosely bound state, which develops into a bound
state as the depth of the potential is further increased.
In the next subsection, we will develop the
formal connection between scattering states, with $a>0$, and bound states.

\subsection{Bound States: $-V_0<E<0$}
We have already briefly introduced the notion of bound states supported by the FSW in any dimension.  
In this section, we will focus our attention on energies for which  $-V_0<E<0$, and thereby make the connection between low-energy {\em bound states} and the low-energy scattering
states discussed above.
In order to accomplish this goal, we must  address how the condition  $-V_0<E<0$ affects our solutions to the TISE in the asymptotically free ($r>b$) 
region; the solutions will be unaffected in the interior region since
$\eta^2 >0$ still remains true.

\subsubsection{Exterior solution: $r>b$}

With $E$ negative, $k=\sqrt{ME/\hbar^2}$ is imaginary, and we can write $k = i\kappa$ where $\kappa \equiv \sqrt{-ME/\hbar^2}$.  We may then use our scattering solution, 
{\it viz.,} Eq.~(11), with $k \to i\kappa$.  The identification $k \to i\kappa$ is only a formal tool, and if desired, a direct solution of Eq.~(10) with $-k^2$ replaced by $\kappa^2$ could
be pursued instead. 

To begin, we make use of the following properties of Bessel functions of imaginary argument:\cite{handbook}
\begin{equation}
J_{\alpha}(i\kappa r) = i^{\alpha}I_{\alpha}(\kappa r)
\end{equation}
and
\begin{equation}
Y_{\alpha}(i\kappa r) = i^{\alpha+1}I_{\alpha}(\kappa r)-\frac{2}{\pi}i^{-\alpha}K_{\alpha}(\kappa r)~,
\end{equation}
where $I_{\alpha}$ and $K_{\alpha}$ are the modified Bessel functions of the first and second kind, respectively.   Utilizing these expressions in Eq.~(11) allows us to write for the exterior
solution, 
\begin{equation}
\psi_{>}(r) = \frac{g_3K_{\alpha}(\kappa r)+g_4I_{\alpha}(\kappa r)}{r^{d/2-1}},
\end{equation} 
where $g_3$ and $g_4$ are dimensionally dependent integration constants.  Of the two modified Bessel functions, $I_{\alpha}$ increases exponentially, and as a result we set $g_4=0$. Thus 
our exterior, normalizable bound state wave function reads\cite{nieto,farrell}
\begin{equation}
\psi_{>}(r) = \frac{g_3K_{\alpha}(\kappa r)}{r^{d/2-1}}.
\end{equation} 

\subsubsection{Interior solution: $r<b$}
As mentioned earlier, the interior solution is again given by Eq.~(9) owing to the fact that for $-V_0 < E<0$, $\eta^2>0$ still holds true.
\subsubsection{Boundary conditions at $r=b$}

The allowed energies of these bound states are determined by matching the wave function and its derivative at the boundary $r=b$. As a result, the energy is determined by solutions to the 
equation
\begin{equation}
\left(\frac{\psi_{<}'(r)}{\psi_{<}(r)}\right)_{r=b} =  \left(\frac{\psi_{>}'(r)}{\psi_{>}(r)}\right)_{r=b}~~,
\end{equation}
which is identical to Eq.~(12), but now applied to bound states.
In 1D, Eq.~(35) reproduces the standard result for the even bound states~\cite{griffiths}
\begin{equation}
\kappa = \eta \tan{\eta b}.
\end{equation}
For $d$-dimensions, Eq.~(35) reads
\begin{equation}\label{boundary}
\kappa\frac{K_{\alpha+1}(\kappa b)}{K_{\alpha}(\kappa b)} = \eta\frac{J_{\alpha+1}(\eta b)}{J_{\alpha}(\eta b)}~. 
\end{equation}
Our intention is to now relate the {\em low energy} bound states to the discussion of the scattering length given in the previous subsection. 
To proceed, we put $k = i \kappa$ in Eq.~(16), use Eqs.~(31) and (32), along with Eq.~(37), to obtain 
the remarkably simple result in {\em all} dimensions:
\begin{equation}
\cot{\delta_0}= i~.
\end{equation}
Therefore, the bound states can be directly obtained from the $E>0$ scattering states in {\em any} dimension provided we put $k\to i\kappa$ and $\cot(\delta_0) = i$.  Recall that for scattering states, the continuity of
the logarithmic derivative of the wave function
at $r=b$ fixed the $s$-wave phase shift.  In the present case, the {\em same} boundary condition at $r=b$ has likewise fixed the phase shift, but now,  the phase shift is a purely imaginary number $\cot(\delta_0)=i$,
associated with the purely imaginary momentum, $k = i\kappa$.  
We wish to point out that in standard treatments of scattering theory, the same result,
{\it viz.,} $\cot(\delta_0)=i$, is obtained, but  involves the introduction of the  $S(k)$-matrix (or equivalently, the partial-wave scattering amplitude), and an analysis of its analytic properties.~\cite{shea2008,taylor}  
Here, we have accomplished the same goal, but without having
to introduce
any additional mathematical machinery.  While our approach may not be as mathematically elegant, it requires less formalism, and is therefore more accessible to students with only a limited exposure to scattering theory.

Using Eq.~(38), we may finally make the connection between the low-energy bound state energy, $E=-\hbar^2 \kappa^2/M$ and the $s$-wave scattering length.  Quite simply, Eqs.~(19) and (24) are evaluated at
$\cot(\delta_0) = i$, giving
\begin{equation}
k = i^{1/(d-2)} \frac{2}{a}\left(\frac{\Gamma(d/2)\Gamma(d/2-1)}{\pi}\right)^{1/(d-2)}~,~~~~d \ne 2~,
\end{equation}
and
\begin{equation}
k = i \frac{2e^{-\gamma}}{a}~,~~~~d=2~.
\end{equation}
Since $\kappa = {\rm Imag}[k]$, we obtain in both 1D and 3D the well known result~\cite{shea2008,farrell}
$\kappa = 1/a$, and  $E=-\hbar^2/Ma^2$ for the shallow bound state energies.  In 2D, we obtain $\kappa = 2e^{-\gamma}/a$ giving
$E = -4\hbar^2e^{-2\gamma}/Ma^2$ for the shallow bound state energy.~\cite{farrell} 

\section{Contact Interaction Limit of the finite square well}

Following the standard 1D treatment of allowing a FSW to go to a zero-range interaction ({\it i.e.,} delta-function) 
we wish to investigate the $b  \to 0$ limit of the above developed results.  For simplicity, our analysis will be formulated in terms of the low-energy bound states, although {\em exactly the same}
results will also hold true for the low-energy scattering states; this is not surprising in view of our discussions up to now.
We will take the $b \to 0 $ limit of the FSW while insisting that the $d$-dimensional area inside the well remain  constant. This area is given by
\begin{equation}\label{area}
\tilde{V}_0 = \int_{\bf All \ Space} V(r)~d\tau = -\frac{\pi^{d/2}b^d}{\Gamma(d/2+1)}V_0~.
\end{equation}
The negative sign in Eq. (\ref{area}) merely reflects the fact that our $d$-dimensional FSW is {\em attractive} and thus $V(r) \leq 0$ while $r \geq 0$.
To keep $\tilde{V}_0$ constant, we require that
\begin{equation}
V_0 = - \frac{\tilde{V}_0\Gamma(d/2+1)}{\pi^{d/2}b^d} >0~,
\end{equation}
illustrating that as $b \to 0$,  $V_0 \to \infty$. We may now write our TISE in terms of $\tilde{V}_0$. We obtain
\begin{equation}\label{preint}
-\frac{\hbar^2}{M}\nabla_d^2 \psi(r) +\frac{\tilde{V}_0\Gamma(d/2+1)}{\pi^{d/2}}\frac{\Theta(b-r)}{b^d} \psi(r) = E \psi(r)~,
\end{equation}
where it is to be understood that, in place of $\psi(r)$ above, we use $\psi_{<}$ for $r<b$ and $\psi_{>}$ for $r>b$. A standard analysis of a one-dimensional attractive delta function interaction centered at the origin\cite{griffiths} involves integrating the TISE in a neighbourhood about the origin from $-\epsilon$ to $\epsilon$, followed by letting $\epsilon \to 0$.   We extend this approach to the case of
arbitrary dimensions in the sense that we now  integrate our above TISE over a hyper-sphere of radius $\epsilon>b$, and then allow $b \to 0$ (this gives us a contact interaction) followed by $\epsilon \to 0$. 
We will focus on the $d\ne 2$ case now, and present the 2D result at the end of this section.

We recall that for the bound states, we have the following solutions:
\begin{equation}
\psi_{<}(r) =\frac{c_1\eta b^{d/2}}{J_{d/2}(\eta b)}\frac{J_{\alpha}(\eta r)}{r^{d/2-1}}~
\end{equation}
and
\begin{equation}
\psi_{>}(r) = \frac{g_3K_{\alpha}(\kappa r)}{r^{d/2-1}}~.
\end{equation}
The Laplacian of  the exterior solution is given by
\begin{equation}\label{1}
 \nabla_d^2 \psi_{<}(r) =\frac{c_1b^{d/2}\eta^3}{J_{d/2}( \eta b)}\frac{J_{d/2+1}(\eta r)}{r^{d/2-1}}- \frac{c_1b^{d/2}\eta^2d}{J_{d/2}( \eta b)}\frac{J_{d/2}(\eta r)}{r^{d/2}}~.
\end{equation}
Integrating the TISE over a hyper-sphere of radius $\epsilon>b$ we obtain
\begin{eqnarray}\label{integralsBS}
-\frac{\hbar^2}{M}\left(\int_{r=0}^{r=b}\nabla_d^2 \psi_{<}(r)d\tau + \int_{r=b}^{r={\epsilon}}\nabla_d^2 \psi_{>}(r)d \tau\right) +\frac{\tilde{V}_0\Gamma(d/2+1)}{\pi^{d/2}}\int_{r=0}^{r=b}\frac{\psi_{<}(r)}{b^d} d \tau  \\
\nonumber =E\left(\int_{r=0}^{r=b}\psi_{<}(r)d \tau + \int_{r=b}^{r={\epsilon}} \psi_{>}(r)d \tau\right)~.
\end{eqnarray} 
When we let $b \to 0$, followed by $\epsilon \to 0$, it is straightforward to see that in this limiting procedure,
\begin{equation}
\int_{r=0}^{r=b} \nabla_d^2\psi_{<}(r)d\tau~,~\int_{r=0}^{r=b}\psi_{<}(r)d\tau~,~\int_{r=b}^{r=\epsilon} \psi_{>}(r)d\tau~,
\end{equation}
will all vanish.

Next, anticipating $b\to 0$, we expand $\psi_{>}(r)$ for small $r$ and obtain (for $d\ne2$)
\begin{equation}\label{BSsmall}
\psi_{>}(r) \simeq \frac{g_3\Gamma(d/2-1)}{2} \left(\frac{2}{\kappa} \right)^{d/2-1} \frac{1}{r^{d-2}} + {\it O}(r^{3-d})~.
\end{equation}
An application of  the $d$-dimensional generalization of the divergence theorem, {\it viz.,}~\cite{ikhdair,baker,griffithsEM}
\begin{equation}
\nabla^2\left(\frac{1}{r^{d-2}}\right) = - \frac{d(d-2)\pi^{d/2}}{\Gamma(d/2+1)} \delta^{(d)}(r),
\end{equation}
gives us
\begin{equation}
\nabla_d^2\psi_{>}(r) \simeq -\frac{g_3\Gamma(d/2-1)d(d-2)\pi^{d/2}}{2\Gamma(d/2+1)} \left(\frac{2}{\kappa} \right)^{d/2-1} \delta^{(d)}(r)~.
\end{equation}
The integral
\begin{equation}
\frac{\tilde{V}_0\Gamma(d/2+1)}{\pi^{d/2}}\int_{r=0}^{r=b}\frac{\psi_{<}(r)}{b^d} d \tau = c_1 d {\tilde V_0}
\end{equation}
is {\em independent} of $b$.  We may then write Eq. (\ref{integralsBS}) as
\begin{equation}
\frac{\hbar^2g_3\Gamma(d/2-1)d(d-2)\pi^{d/2}}{2M\Gamma(d/2+1)}\left(\frac{2}{\kappa}\right)^{d/2-1} + c_1d\tilde{V}_0 = 0~,
\end{equation}
from which we find
\begin{equation}
\tilde{V}_0 = - \frac{g_3}{c_1} \frac{\hbar^2 \Gamma(d/2-1)}{2M\Gamma(d/2+1)}(d-2)\pi^{d/2}\left(\frac{2}{\kappa}\right)^{d/2-1}.
\end{equation}
This form for $\tilde{V}_0$ ensures that our TISE has the proper behaviour under integration in the $b\to 0$ limit.
This expression for $\tilde{V}_0$ gives us a potential that reads
\begin{equation}
V(r) = - \frac{g_3}{c_1} \frac{\hbar^2 \Gamma(d/2-1)}{2M}(d-2)\left(\frac{2}{\kappa}\right)^{d/2-1}\lim_{b\to 0}\frac{\Theta(b-r)}{b^d} ~.
\end{equation}
The issue now is that the potential in Eq.~(55) only acts on $\psi_{<}(r)$, but in the zero-range limit, $\psi_{<}(r)$ will be ``squeezed out'' as the
entire interior region is reduced to a point.
Mathematically, the remedy to this situation is to insist that
\begin{equation}
\int \lim_{b\to0} \frac{\Theta(b-r)}{b^d}\psi_{<}(r) d \tau =  \int \Lambda_d \delta^{(d)}(r)\hat{O}^{(d)}\psi_{>}(r) d \tau~,
\end{equation}
where $\hat{O}^{(d)}$ is an operator to be determined, and $\Lambda_d$ is a dimensionally dependent constant. $\hat{O}^{(d)}$ is sometimes referred to as a regularization operator~,\cite{wodkiewicz}
and its neccessity can be traced back to the fact that $\forall d \ne 1$, $\psi_{>}(r)$ is undefined as $r\to 0$.  Thus, if we were to try to use a bare delta function in the integral
on the right-hand side of Eq.~(56), the integral would be ill-defined $\forall d \ne 1$.

The integral on the left-hand side of Eq.~(56) is
\begin{equation}\label{LHSBS}
\int \lim_{b\to0} \frac{\Theta(b-r)}{b^d}\psi_{<}(r) d \tau= \frac{c_1d\pi^{d/2}}{\Gamma(d/2+1)}~,
\end{equation}
whereas for the right-hand side we have
\begin{equation}\label{RHSBS}
\int \Lambda_d \delta^{(d)}(\rv)\hat{O}^{(d)}\psi_{>}(r) d \tau = \Lambda_d g_3 \lim_{r\to0} \hat{O}^{(d)}\left(\frac{K_{\alpha}(\kappa r)}{r^{d/2-1}}\right)~.
\end{equation}
Equating the results of Eqs. (\ref{LHSBS}) and (\ref{RHSBS}), we obtain the constant $\Lambda_d$, {\it viz.,}
\begin{equation}
\Lambda_d = \frac{c_1}{g_3}\frac{d\pi^{d/2}}{\Gamma(d/2+1)}\frac{1}{ \lim_{r\to0} \hat{O}^{(d)}\left(\frac{K_{\alpha}(\kappa r)}{r^{d/2-1}}\right)}~.
\end{equation}
We are now free to replace $\lim_{b\to0} \frac{\Theta(b-r)}{b^d}$ with $\Lambda_d \delta^{(d)}(r)\hat{O}^{(d)}$ in Eq.~(55) to get, for the $b\to 0$ limit of the FSW,
\begin{equation}
V(r) = -  \frac{\hbar^2d\pi^{d/2} \Gamma(d/2-1)}{2M\Gamma(d/2+1)}(d-2)\left(\frac{2}{\kappa}\right)^{d/2-1}\frac{\delta^{(d)}(r)}{ \lim_{r\to0} \hat{O}^{(d)}\left(\frac{K_{\alpha}(\kappa r)}{r^{d/2-1}}\right)}\hat{O}^{(d)}.
\end{equation}
The appropriate operator, $\hat{O}^{(d)}$, has different forms depending on the dimensionality of the system. 
Specifically, $\hat{O}^{(d)}$ is found by requiring that the singular behaviour of $\lim_{r \to 0} \psi_{>}(r)$ in Eq.~(56) is removed.  
Let us orient ourselves first with 1D, and then move on to higher dimensional spaces.  

In 1D, $\psi_{>}(r)$ is regular as $r\to 0$, so  the proper operator is $\hat{O}^{(1)} = 1$ and we have
\begin{equation}
V(r) =   -\frac{\hbar^2\sqrt{2\pi\kappa}}{M}\frac{\delta^{(1)}(r)}{ \lim_{r\to0}\left(\sqrt{r}K_{-1/2}(\kappa r)\right)}~.
\end{equation}
We note that $\lim_{r\to0}\left(\sqrt{r}K_{-1/2}(\kappa r)\right) = \sqrt{\pi/2\kappa},$ and we obtain the known result
\begin{equation}
V(r) =   -\frac{2\hbar^2\kappa}{M}\delta^{(1)}(r) =  -\frac{2\hbar^2}{Ma }\delta^{(1)}(r)~,
\end{equation}
where we have made use of $\kappa=1/a$ for the shallow bound state found in Sec.~III B above.  Equation (62) is to be viewed as the zero-range interaction
reproducing the same $k \to 0$ scattering as for {\em any} short-range potential with the same scattering length, $a$.  Again, note that in 1D, the $b\to 0$ limit of the
FSW is proportional to a {\em bare} delta function potential.

For  higher dimensions ($d \ne 2$), the operator needed is $\hat{O}^{(d)} = \left(\frac{\partial}{\partial r}\right)^{d-2} r^{d-2}$, with
\begin{equation}
\lim_{r \to 0}\left(\frac{\partial}{\partial r}\right)^{d-2} r^{d-2} \frac{K_{\alpha}(\kappa r)}{r^{d/2-1}} = \frac{(-1)^{\frac{d-1}{2}}\Gamma(d-1)\pi}{2\Gamma(d/2)}\left(\frac{\kappa}{2}\right)^{d/2-1}~.
\end{equation}
In defining the operator $\hat{O}^{(d)}$, we have kept to the common convention in the literature and used the partial derivative, $\partial/\partial r$, to emphasize that $\hat{O}^{(d)}$ only acts
on the radial component of any function it encounters. Of course, if
$\hat{O}^{(d)}$ acts on a function of $r$ only, the partial derivative is to be treated as a full derivative.

The expression for the $b\to 0$ limit of the FSW then becomes
\begin{equation}
V(r) = \frac{\hbar^2 d \pi^{d/2}}{M\Gamma(d/2+1)\Gamma(d-2)} \left[ -\frac{(-1)^{\frac{d-1}{2}}\Gamma(d/2)\Gamma(d/2-1)}{\pi}\left(\frac{2}{\kappa}\right)^{d-2}\right] \delta^{(d)}(r) \left(\frac{\partial}{\partial r}\right)^{d-2} r^{d-2}~.
\end{equation}
For example, setting $d=3$ in Eq.~(64) gives
\begin{eqnarray}
V(r) &=& \frac{4\pi \hbar^2}{M\kappa} \delta^{(3)}(r) \frac{\partial}{\partial r} r \nonumber \\ 
&=& \frac{4\pi \hbar^2 a}{M} \delta^{(3)}(r) \frac{\partial}{\partial r} r~,
\end{eqnarray}
where again, for  low-energy bound states, $\kappa = 1/a$.  Equation (65) is in perfect agreement with what is found using other approaches in the literature.~\cite{shea2008,reg3dnote,wodkiewicz, li}

For 2D, we need to step a little bit further back. The small $r$ expression given in Eq. (\ref{BSsmall}) is valid for all $d \ne 2$. For $d=2$ we have
\begin{equation}
\psi_{>}(r) \simeq - g_3\left(\ln{\kappa r/2}+\gamma\right)~,
\end{equation}
with 
\begin{equation}
\nabla_2^2\psi_{>}(r) = -2\pi g_3 \delta^{(2)}(r)~.
\end{equation}
Except for the term involving $\nabla_2^2\psi_{>}(r)$, all of the integrals in Eq. (\ref{integralsBS}) are the same in 2D as they are in any other dimension. As a result, our integrated TISE for 2D is
\begin{equation}
\frac{\pi 2g_3\hbar^2}{M} + 2c_1\tilde{V}_0 = 0~,
\end{equation}
and we obtain
\begin{equation}
\tilde{V}_0 = - \frac{g_3}{c_1}\frac{\pi \hbar^2}{M}.
\end{equation}
So in 2D we have
\begin{equation}
V(r) = - \frac{g_3}{c_1}\frac{\hbar^2}{M}\lim_{b \to 0}\frac{\Theta(b-r)}{b^2}.
\end{equation}
Again, we wish to replace the part involving the Heaviside function with an operator involving the delta function. By the same argument as for all other dimensions, we obtain
\begin{equation}
\Lambda_2 = \frac{c_1}{g_3}\frac{2\pi}{ \lim_{r\to0} \hat{O}^{(2)}\left(K_{0}(\kappa r)\right)}~,
\end{equation}
and our 2D zero-range potential becomes
\begin{equation}
V(r) = -\frac{2\pi\hbar^2}{M} \frac{\delta^{(2)}(\rv)}{ \lim_{r\to0} \hat{O}^{(2)}\left(K_{0}(\kappa r)\right)} \hat{O}^{(2)}.
\end{equation}
The proper operator for 2D is $\hat{O}^{(2)} = 1-r\ln{\frac{\kappa r}{2e^{1-\gamma}}}\frac{\partial}{\partial r}$, and we note that
\begin{equation}
 \lim_{r\to0} \left(1-r\ln{\frac{\kappa r}{2e^{1-\gamma}}}\frac{\partial}{\partial r}\right) K_0(\kappa r) = -1~.
 \end{equation}
 It follows that the zero-range limit of the FSW in 2D is given by
 \begin{eqnarray}
 V(r) &=& \frac{2\pi \hbar^2}{M} \delta^{(2)}(r)  \left(1-r\ln{\frac{\kappa r}{2e^{1-\gamma}}}\frac{\partial}{\partial r}\right) \nonumber \\
 &=& \frac{2\pi \hbar^2}{M} \delta^{(2)}(r)  \left(1-r\ln{\frac{r}{a e}}\frac{\partial}{\partial r}\right)~,
 \end{eqnarray}
where Eq.~(40) has been used to connect the low-energy bound state to a scattering state with $a>0$.

\section{conclusions}

We have presented a systematic analysis of the FSW in arbitrary dimensions, thereby providing generalizations to quantities typically introduced in the context of 3D scattering; namely, the $s$-wave
scattering length and phase shift~.\cite{taylor}  We have shown that the $d$-dimensional scattering length, $a$, can be naturally interpreted as the node (or its extrapolation) of the $k \to 0$ wave function.  In the
$k \to 0$ limit, we have also illustrated how the $s$-wave scattering length completely characterizes the scattering properties of the system for {\em any} short-range potential in any dimension.  For students familiar
familiar with 3D scattering, the 2D results of the FSW highlight that even in simple ``toy-model potentials'', dimensionality plays a pivotal role in determining the physical properties of the system.  For example,
we point out that in 1D and 2D, $a>0$ for any $V_0$, which introduces the notion that an arbitrarily weak short-range attractive interaction in 1D and 2D will always support at least one bound state~.\cite{chandan}

In order to avoid introducing additional mathematical formalism, we have also shown how to utilize {\em only} the scattering states with imaginary momentum, $i\kappa$, to obtain imaginary
phase shifts, which for a bound state near $E \to 0^-$, can be connected to the $s$-wave scattering length discussed earlier.  This alternative approach has the virtue of utilizing  mathematical
techniques familiar from the study of elementary 1D problems
{\it i.e.,} continuity of the wave function and its deriative at the boundary to treat both bound and scattering states in a unified way.  We feel that students may benefit from this presentation, 
particularly those not yet exposed to the analytic properties of the $S(k)$-matrix~.\cite{taylor,shea2008, farrell}

Following our treatment of the $d$-dimensional FSW, we examined its $b\to 0$ limit by extending the 1D analysis~\cite{griffiths} to arbitrary dimensions.  Our main result is that the
zero-range limit of the FSW is given by
\begin{equation}
V(r) = -  \frac{\hbar^2d\pi^{d/2} \Gamma(d/2-1)}{2M\Gamma(d/2+1)}(d-2)\left(\frac{2}{\kappa}\right)^{d/2-1}\frac{\delta^{(d)}(r)}{ \lim_{r\to0} \hat{O}^{(d)}\left(\frac{K_{\alpha}(\kappa r)}{r^{d/2-1}}\right)}\hat{O}^{(d)}~,
\end{equation}
for $d \ne 2$, and 
\begin{eqnarray}
 V(r) &=& \frac{2\pi \hbar^2}{M} \delta^{(2)}(r)  \left(1-r\ln{\frac{\kappa r}{2e^{1-\gamma}}}\frac{\partial}{\partial r}\right)~,
 \end{eqnarray}
 for $d=2$, respectively.  The non-trivial expressions for the $b\to 0$ limit of the FSW serve to illustrate that the usual textbook suggestion of obtaining the delta function results from the limit
 of a FSW is not so straightforward in dimensions greater than one.   In fact, the $d=2$ case turns out to be the most interesting, with subtle mathematical issues not typically
 discussed in undergraduate quantum mechanics courses.  Nevertheless, we have tried to use only the TISE and a minimal amount of mathematical machinery to
 motivate the so-called regularized delta function potentials given by Equations (75) and (76).  In more technical papers, similar results are developed in the context of self-adjoint
 extensions, and Green's function techniques and are given the name ``pseudo-potentials''.~\cite{shea2008, farrell, wodkiewicz, li, olshanii1, kanjilal}   
 
 Finally, it is our hope that the presentation used in this paper may serve as a basis for introducing a more general treatment of the FSW in undergraduate quantum mechanics.   To this end, we
 suggest the following useful exercise.  Initially, the student would be asked to explore Eq.~(37) using standard graphical solutions, for $d=1$ and $d=3$, with {\em fixed} $b$, and varying $V_0$.  Next the student would be
 asked to fix $V_0b^d$ and let $V_0 \to \infty$ and $b\to0$ and also graphically look for solutions.  The student will find that in the latter limit, the 1D version of Eq.~(37)  has only one, finite energy, bound state, while
 the 3D solution obtains more and more bound states as $V_0$ increases.  The point of this exercise would be to show that a naive replacement of the 1D FSW with a delta function potential leads to only one bound
 state in 1D, which is perfectly consistent with known results.~\cite{calogero}  However, the analogous $V_0\to \infty,~b \to 0$ limit in 3D (with $V_0b^3$ fixed) leads to an infinite number of bound states, with an unphysical, infinitely bound
 ground state energy.  These findings would then motivate
 a different approach to the zero-range limit of the FSW in 3D, namely,  the approach suggested in this paper.
 The student could then be guided to obtain the appropriately regularized 3D delta potential, and if desired, the connection between the regularized 3D delta potential and self-adjoint Hamiltonians\cite{fulop,araujo}
 could be pursued.
 
 \acknowledgments
 We would like to thank Dr. R. K. Bhaduri and Dr. C. Hanna for useful discussions.  This work was supported by
 the National Sciences and Engineering Research Council (NSERC) of Canada through the Discovery Grant program.
 A. Farrell would also like to acknowledge the NSERC USRA program for additional financial support.
 


\begin{thebibliography}{99}
\bibitem{note2}
Retaining the next leading order term in the expansion of Eq.~(16) about small $k$ would lead us to the
effective range expansion, with the next leading order term proportional to $r_0$, the effective range.
As we will ultimately be interested only in the limit $b \to 0$, and since $r_0 \to 0$ in this limit, we
do not consider the additional terms in the expansion of Equation (16).
\bibitem{note3}
In the language of Ref. [\onlinecite{barlette}] we have set $L=0$, which results in even functions, corresponding to the analogous 1D $s$-wave solutions.
\end{thebibliography}
\end{document}